\documentclass[fleqn,10pt]{wlscirep}

\usepackage{graphicx}
\usepackage{hyperref}
\usepackage{lipsum}
\usepackage{amsmath}
\usepackage{color}

\newcommand{\BigO}[1]{\ensuremath{\operatorname{O}\bigl(#1\bigr)}}

\usepackage{overpic}

\usepackage[normalem]{ulem}

\begin{document}

\author[1]{Dijana Toli{\'c}}
\author[2]{Kaj-Kolja Kleineberg}
\author[2,*]{Nino Antulov-Fantulin}
\affil[1]{Laboratory for Machine Learning and Knowledge Representations, Rudjer Bo{\v s}kovi\'c Institute, Zagreb, Croatia}
\affil[2]{Computational Social Science, ETH Z\"urich, Clausiusstra{\ss}e 50, 8092 Z\"urich, Switzerland}

\affil[*]{anino@ethz.ch}

\title{Simulating SIR processes on networks using weighted shortest paths}

\date{\today}
\begin{abstract}
We present a framework to simulate SIR processes on networks using weighted shortest paths. 
Our framework maps the SIR dynamics to weights assigned to the edges of the network, which can be done for Markovian and non-Markovian processes alike. 
The weights represent the propagation time between the adjacent nodes for a particular realization.  
We simulate the dynamics by constructing an ensemble of such realizations, which can be done by using a Markov Chain Monte Carlo method or by direct sampling. The former provides a runtime advantage when realizations from all possible sources are computed as the weighted shortest paths can be re-calculated more efficiently. 
%
%
We apply our framework to three empirical networks and analyze the expected propagation time between all pairs of nodes. 
Furthermore, we have employed our framework to perform efficient source detection and  to improve strategies for time-critical vaccination. 

\end{abstract}

\maketitle

\section*{Introduction}

Modern socio-technological systems can be described as networks with naturally occurring spreading phenomena such as information propagation in social networks~\cite{Guille2013,lognormalSpreading}, or the epidemic spread among individuals through their contact network ~\cite{anderson:spreading, Vespignani2011, RevModPhys.87.925, Boccaletti2006175}. 
In order to exactly solve the spreading process one must be able to evaluate the probabilistic evolution of the process configurations in time i.e. solve a set of coupled differential equations called master equation \cite{VespginaniBook, Gillespie1992_MA, VanMieghemMC, VanMieghemMC2} or Chapman-Kolmogorov equation.
However, due to the exponential growth of possible configurations only in small number of special cases \cite{VanMieghemMC, VanMieghemMC2} it is possible to solve the master equation directly.
Therefore, different approximations are used: 
(i) neglecting dynamical correlations by assuming statistical independence at first neighbourhood, pair or node level in mean field approximations~\cite{Gleeson,Castellano10, EpidScaleFree, Sharkey2011, Sharkey2015, Kiss2015},
(ii) neglecting loops in the network structure with message passing or belief propagation techniques~\cite{Karrer2010, DMP_0, BBP},
or
(iii) neglecting the temporal evolution by mapping to the percolation process \cite{mollison1977spatial, GrassbergerPercolation, ContEpd2}.
Alternative way is the simulation or sampling the statically correct process configurations or realizations with the dynamic Monte Carlo techniques \cite{Gillespie, dynamicMC, kinteicMC, BogunaNonMarkovian, TemporalNetGillespie, FastSIR}.




In this paper, we propose how to exactly transform the SIR spreading dynamics to a shortest path problem on an ensemble of weighted networks using a Monte Carlo approach. In this framework, weights represent interaction time delays along edges. 
We show the equivalence between the time respecting paths (shortest paths) on the constructed weighted networks and the propagation times of spreading dynamics.  
The mapping is applicable to the generalized Susceptible Infected Recovered (SIR) model~\cite{RevModPhys.87.925} without memory (exponential inter-event distribution/Poisson process) and with memory (arbitrary inter-event distributions).
Contrary to the standard dynamic Monte Carlo methods\cite{Gillespie, dynamicMC, kinteicMC, BogunaNonMarkovian, TemporalNetGillespie, FastSIR}, we do not have to specify initial conditions upfront and we can sample new realizations from the previous ones by making local random Markov Chain transitions between the weighted networks.
In a limit of infinite process time, we establish the connection with bond percolation theory~\cite{mollison1977spatial, GrassbergerPercolation,NewmanPercolation, EpiPercolationNet}.
In the special case of independently assigned edge weights, we establish the connection to the 
disordered in networks~\cite{ShlomoDiorderPRL1, ShlomoDiorderPRL2, PhysRevE.70.046133, VanMieghemDisorder1, VanMieghemDisorder2}.
Finally, we show the applications to source detection \cite{AntulovFantulin2015, Zaman1, pinto2012, DMP_0, BBP, Brockmann} and time-critical vaccination \cite{Jalili2017, Wang2016, Helbing2015}.

\section*{Methods}

\subsection*{Mapping SIR dynamics to weighted shortest paths}

\begin{figure*}[t]
\centering
 \includegraphics[width=0.8\linewidth]{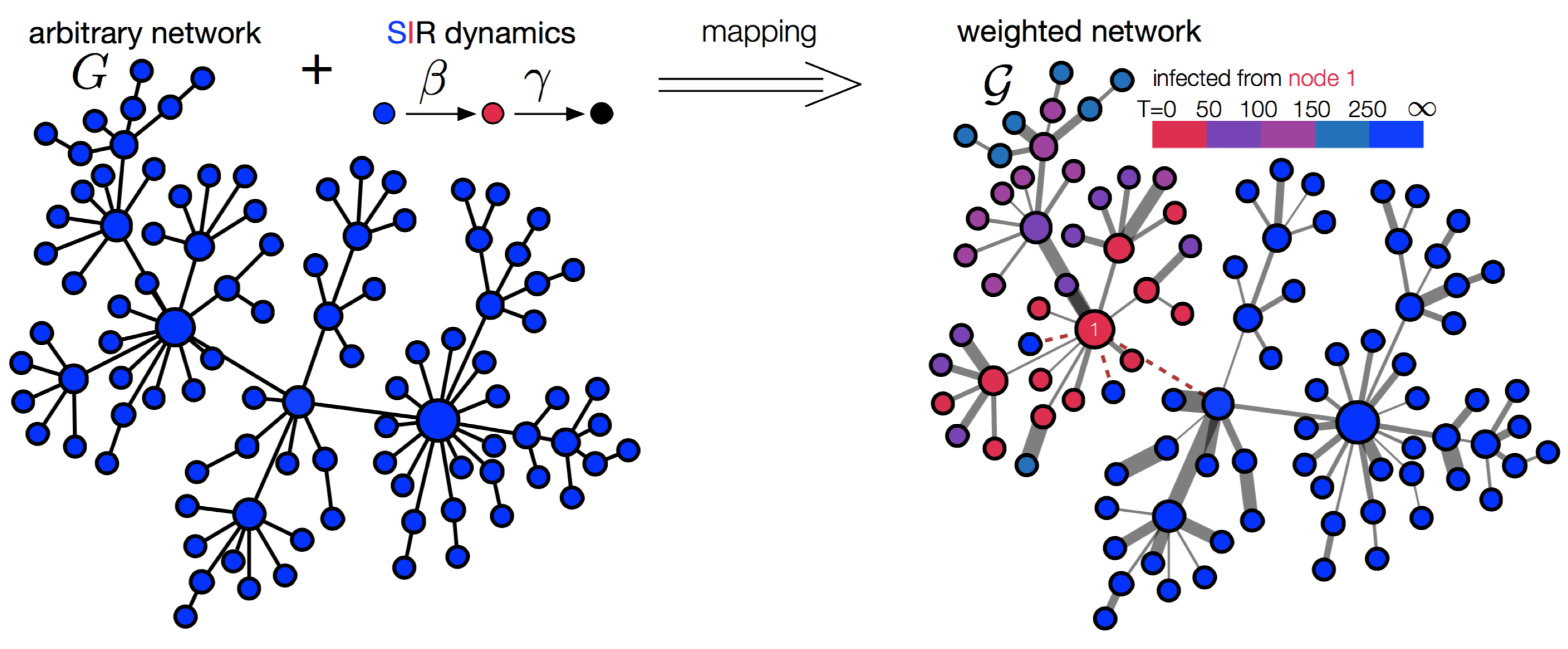}
 \caption{\textbf{The mapping of spreading dynamics to an ensemble of weighted networks}. The weights represent the propagation time delays. Each weighted network represents one realization of the stochastic dynamics from an arbitrary source node. 
The propagation time from the source node to any other node is equal to the shortest path length in weighted network (color coded). The dotted line represents edges with infinite edge weight.
 \label{fig1}}
\end{figure*}


Let us start with the Poissonian SIR model, which is memoryless, such that the next state only depends on the present state of the system. This process is defined by two parameters $(\beta,\gamma)$, which describe 
transitions where susceptible nodes become infected with rate $\beta$ from infected neighbours and transitions where infected nodes recover with rate $\gamma$. 
This process has exponential inter-event time distributions: $\psi(t)=\beta e^{-\beta t}$ (spread) and $\phi(t)=\gamma e^{-\gamma t}$ (recovery). 
However, many realistic processes have memory \cite{BogunaNonMarkovian} and their inter-event time distributions are non-exponential. Therefore, we allow for arbitrary inter-event density distributions $(\psi(t), \phi(t))$ for generalized SIR processes. 

For a given network $G$ and class of generalized SIR spreading models, we
show how to create weighted networks $\left\lbrace \mathcal{G}_k \right\rbrace $ which encode realizations of the stochastic spreading dynamics. Each weighted network $\mathcal{G}_k$ (see Fig.~\ref{fig1}) encodes one possible outcome of the spreading process from every potential source node in a network. 
In particular, a
\textit{time-respecting weighted network} instance $\mathcal{G}_k$ is created by taking the input network $G$ and assigning 
weights to the edges of the network instance with the Inverse Smirnov transform \cite{DevroyeInverseSampling} as
 \begin{equation}
\label{eq:weights}
\rho_{i,j} =
\begin{cases} 
      \Psi^{-1}(x): & \Psi^{-1}(x)\leq \Phi^{-1}(y)\\
      \infty: & \Psi^{-1}(x) > \Phi^{-1}(y) \\
   \end{cases},
\end{equation}
where $x$ and $y$ are uniform random numbers $\in [0,1]$, $\Phi^{-1}(x)$ and $\Psi^{-1}(y)$ are inverse functions of the cumulative inter-event distributions $\Phi(t) = \int_0^t dt' \phi(t')$ and $\Psi(t) = \int_0^t dt' \psi(t')$. 
The quantities $\Psi^{-1}(y)$ and $\Phi^{-1}(x)$ respectively represent the samples of the transmission and recovery time obtained with the Inverse Smirnov transform 
of inter-event distributions. 
In the special case of the Poissonian SIR process one obtains
\begin{equation}
\label{eq:Poissonweights}
\rho_{i,j} =
\begin{cases} 
      -\ln(x)/\beta: & -\ln(x)/\beta \leq -\ln(y)/\gamma \\
      \infty: & -\ln(x)/\beta > -\ln(y)/\gamma \\
   \end{cases}.
\end{equation}
If nodes recover faster than a transmission occurs, the weight is set to infinity to indicate no transmission through the edge.

We denote the distance as shortest paths on weighted networks, $\mathit{d}_{\mathcal{G}_k} (v_i,v_j) = \min_{\chi_{ij}} \sum_{(k,l)\in\chi_{ij}} \rho_{k,l}$, where $\chi_{ij}$ is the set of all possible paths from node $v_i$ to node $v_j$ on network $\mathcal{G}_k$ and $\rho_{k,l}$ denotes the weights defined in Eq.~\eqref{eq:weights}. 
Importantly, this distance is equivalent to the propagation time from node $i$ to $j$, i.e. $t(v_i \rightarrow v_j) = \mathit{d}_{\mathcal{G}_k} (v_i,v_j)$ (\textit{time respecting path equivalence}, see Supplementary information, Sec.~1). The run-time complexity of finding shortest paths \cite{Dijkstra1959} from specific source node to others is $O(E+N\log N)$, where $N$ denotes the number of nodes and $E$ number of edges in a network.
The \textbf{exact mapping} is done by generating a random variable $y$ for each node and variable $x$ for each edge.
This takes all dynamical correlations into the account and generates the ensemble of directed weighted networks. A \textbf{simplified mean-field mapping} can be obtained by generating random variables $x$ and $y$ per each edge, independently. This simplified mean-field mapping holds when $\beta >> \gamma$ (SI case) and generates the ensemble of undirected weighted networks (see Supplementary information, Sec.~5 and 6. for more details). In the section Additional information of this paper, we provide a link to the code that implements the proposed mapping. \\

\begin{figure*}[t]
\centering
 \includegraphics[width=0.8\linewidth]{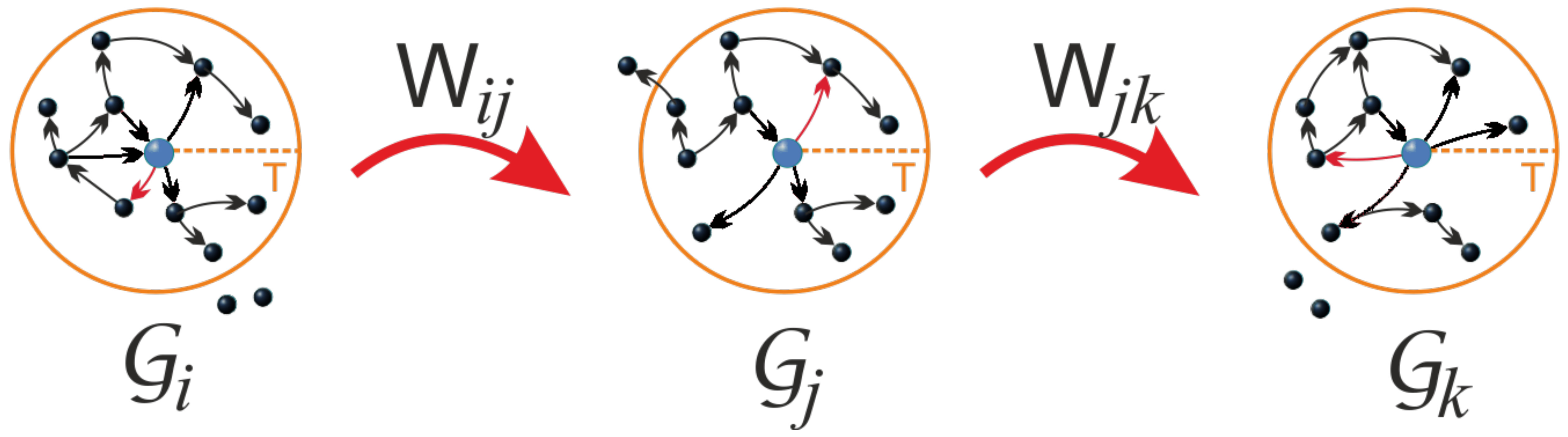}
 \caption{ \textbf{Markov Chain Monte Carlo sampling.} The transitions $W_{i,j} =
w(G_i \rightarrow G_j )$ between different instances of weighted networks in the ensemble.
Randomization of the weight of an edge (denoted by red) corresponds to the transition between the states (realizations of the process). 
 \label{fig2}}
 \end{figure*}

\begin{figure}[t]
\begin{center}
\includegraphics[width=0.75\linewidth]{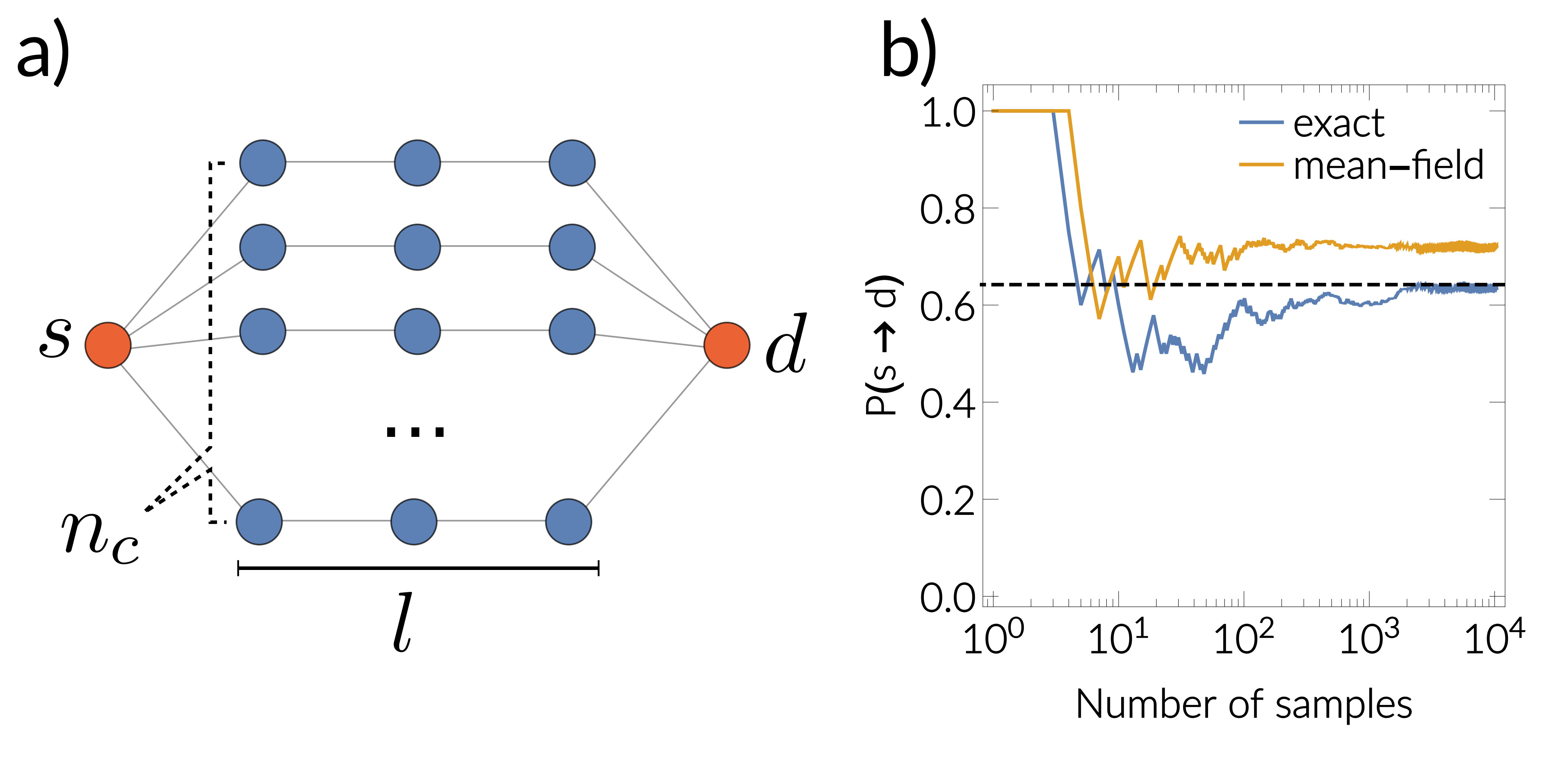}
\end{center}
\caption{ \textbf{Toy network model example}.
\textbf{(a)} Sketch of simple toy network, that consists of $n_c$ chains with $l$ nodes.
The probability that source node $s$ infects destination node $d$ can be calculated analytically (see supplementary information, Sec. 5. for details) and is given by $P(s \rightarrow d) =  1 - \sum_{j=0}^{n_c}  p_{n,j} \cdot \left(1 - p_{1,1}^l \right)^j$. The term $p_{n,k}$, denotes transmissibility for the first neighborhood Eq. 7 for Poissionan SIR process and Eq. 6 for generalized non-Markovian process. \textbf{(b)} Simulation results on the toy network with $n=20$, $l=3$ with Poisson SIR $(\beta=1, \gamma=1)$. The blue curve represents the estimations with the exact mapping (conditional independence among edge weights), the red curve represents the mean-field mapping  (weights are independent) and the black dotted line represents the analytical solution (see supplementary information, Sec. 5. for a million size toy network experiment).} 
\label{fig:pathsAnalytical}
\end{figure}

\subsection*{Sampling methods for simulation}
The instances of the ensemble of weighted networks can be sampled either (i) independently, 
or (ii) by traversing elements of the ensemble with Markov Chain. 

The Markov Chain (rejection-free Gibbs sampler -- see Supplementary information, Sec.~3)
consists of
transitions $W_{i,j} = w(\mathcal{G}_i \rightarrow \mathcal{G}_j)$ over the set of weighted networks $\left\lbrace \mathcal{G}_k \right\rbrace $, where each weighted network corresponds to one realization
of the spreading process from every potential source node. 
This allows us to
construct transitions between weighted networks by changing the weights in the first neighbourhood of a randomly selected node
for exact mapping or by assigning a new weight to the randomly selected edge for a mean-field mapping, both according to Eq.~\eqref{eq:weights} (see Fig.~\ref{fig2}).
The existence of a stationary distribution of this Markov Chain is guaranteed by detailed balance property and the uniqueness by ergodicity (see Supplementary information, Sec.~3).
In other words, the Gibbs sampling---as well as independent sampling---constructs the probability distribution $P(\mathcal{G})$ of the weighted network ensemble, such that it is a statistically exact representation of the stochastic process. 
From the ensemble of weighted networks, due to ergodicity, the expectation $\left\langle f(\mathcal{G}) \right\rangle $ can be estimated as the average over $n$ samples of weighted networks:
\begin{equation}
\left\langle f(\mathcal{G}) \right\rangle = \sum_{\mathcal{G}_k} P(\mathcal{G}_k) f(\mathcal{G}_k) \approx 1/n \sum_{k=1}^n f(\mathcal{G}_k)
\label{eq:estimation}
\end{equation}
By changing functions $f(\mathcal{G})$, we estimate different properties such as total outbreak size, temporal evolution, expected propagation time or source likelihood (see Supplementary information, Sec.~4). 
The same estimation formula \ref{eq:estimation} can be applied to the independent sampling. The convergence rate of independent estimation are bounded with the Berry–Esseen inequality as $O(n^{-1/2})$.
If all to all shortest paths need to be calculated, after each weight change in Markov Chain transition, shortest paths can be dynamically recalculated~\cite{DSP,DSP_APSP} with computational complexity $\BigO{N^2 log^3 N}$, where $N$ denotes the number of nodes in a network (see Supplementary information, Sec.~8 for more details). 
In Fig.~\ref{fig:pathsAnalytical}, we demonstrate that the estimations converge to the analytical solution on a toy network model. This confirms that the shortest paths in the ensemble of weighted networks are taking into the account stochastic spreading along all possible paths in the original unweighted network. 

\subsection*{Relation to bond percolation}

\begin{figure}[h]
\begin{center}
\includegraphics[width=0.65\linewidth]{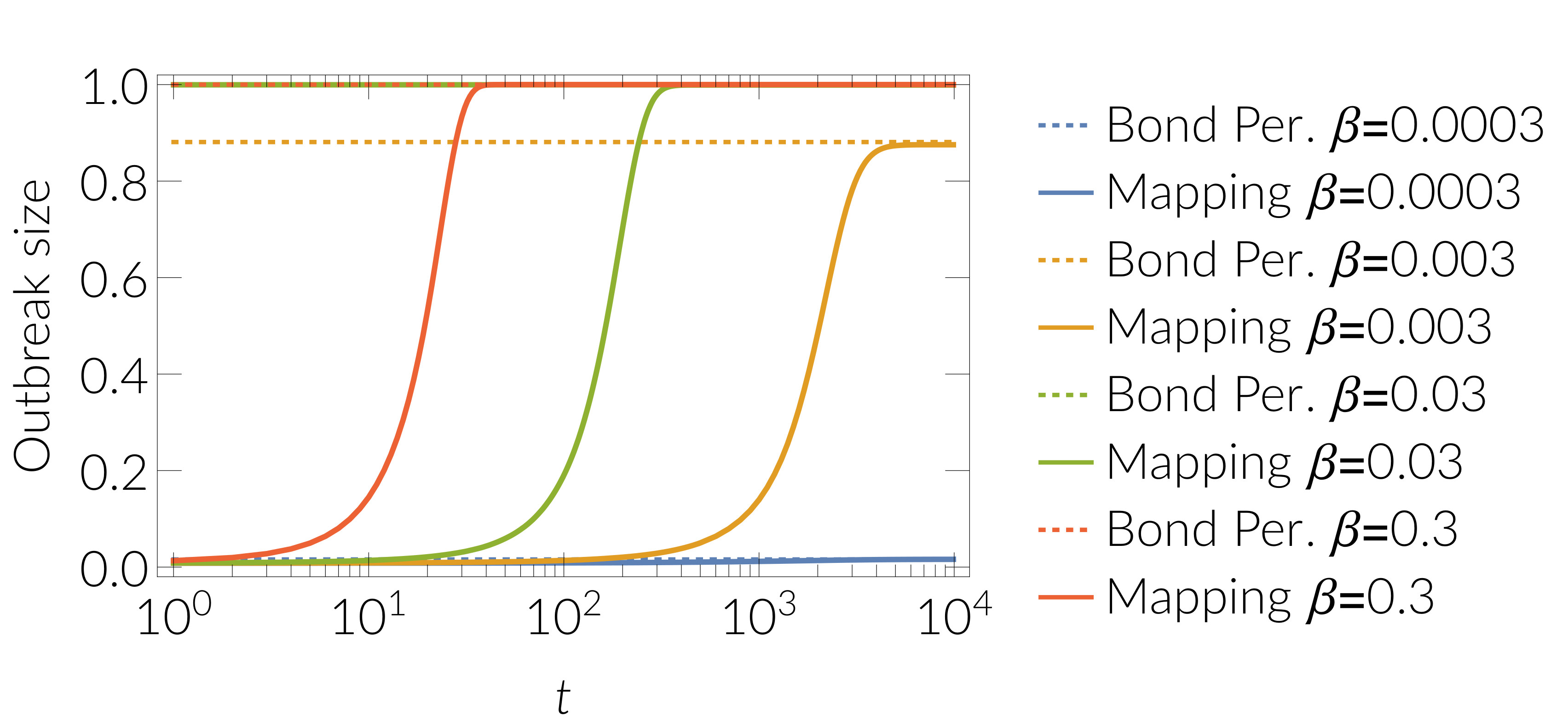}
\end{center}
\caption{\textbf{Equivalence to bond percolation.}
Average outbreak size in time estimated with the proposed mapping and comparison with bond percolation late-time outbreak size (dotted lines), see Eq.\ref{eq:limitFormula}. In a limit of infinite time the mapping becomes equivalent to the bond percolation. Results were obtained for the SIR dynamics (continuous time) with different transmission rates $\beta$ (0.3, 0.03, 0.003, 0.0003) and recovery rate $\gamma = 0.001$ and with $n=10^4$ simulations on the 4-connected two dimensional regular lattice (11x11).}
\label{fig:bond}
\end{figure}

In the following, we state the connection to percolation theory. 
From the perspective of source node $v_i$, the shortest path $\mathit{d}_{\mathcal{G}}(v_i, v_j)$ indicates the first infection time of node $v_j$. Therefore all nodes that are not reachable within the temporal distance $t$ stay in the susceptible state. If a node is reachable within the temporal distance $t$, then it was infected.
By letting $t$ go to infinity we obtain asymptotic realization 
\textit{equivalence to bond percolation}. In particular, the size of the set of nodes $v_j$ that are reachable in any finite time from the source node $v_i$, $|S_p (v_i)|$, is given by 
\begin{equation}
 |S_p (v_i)| = \lim_{t \to \infty}  | \{ \mathit{d} (v_i,v_j) < t\} |.
 \label{eq:limitFormula}
\end{equation}  
The full proof is given in the Supplementary information, Sec.~2. and holds for simplified mean-field mapping, where edge weights are independent. 

Hence, $|S_p (v_i)|$ denotes the size of the connected bond percolation component,
where the transmissibility parameter~\cite{NewmanPercolation} from source $v_i$ is given by \begin{equation}
p =  \int_0^\infty d\tau \phi_i(\tau) \int_0^\tau d\tau'\psi_i(\tau').
\end{equation}  
In Fig.~\ref{fig:bond}, we present the experiments where the equivalence to bond percolation is validated.
The transmissibility quantifies the probability that an infected node transmits infection along a link to a
susceptible neighbour before it recovers. The second integral accounts for the conditional probability of transmission up to the fixed recovery time $\tau$, and the first integral account for all possible recovery times  $\tau$. 
Furthermore, to take dynamical correlations into the account that were previously missing in~\cite{NewmanPercolation}, we generalize and formalize the transmissibility for the first neighborhood or, more precisely, the probability $p_{n,k}$ that $k$ out of $n$ directed links will be active
\begin{equation}
p_{n,k} = \int_0^\infty \phi(\tau) d\tau \binom{n}{k} \left( 1-\Psi(\tau) \right)^{(n-k)} \Psi(\tau)^k,
\label{eqn:generalPnk}
\end{equation} 
where $\Psi(\tau) = \int_0^\tau \psi(t) dt$. 
For Poissionan process this becomes
\begin{equation}
p_{n,k} = \binom{n}{k} \frac{\gamma}{\beta}\frac{ \Gamma(k+1)\Gamma(\frac{\gamma+\beta(n-k) }{\beta})}{ \Gamma(k+1+\frac{\gamma+\beta(n-k) }{\beta})} .
\label{eqn:poissionPnk}
\end{equation}

The generalized transmissibility for the first neighborhood Eq.\ref{eqn:poissionPnk} establishes the connection to semi-directed bond percolation~\cite{EpiPercolationNet} (see Supplementary information, Sec.~5 and 6). To have exact mapping of the SIR process to percolation, having a single percolation parameter $p$ is no longer accurate when $\beta \approx \gamma$ and generalized transmissibility $p_{n,k}$ for the first neighborhood is needed to capture all dynamical correlations.

\subsection*{Relation to disordered networks}


The question how the average propagation time between two individuals scales with the system size is crucial for many real applications and of special theoretic interest. 
In particular, if this distance scales logarithmically with the system size, an epidemic can spread nearly instantaneously in arbitrary large networks. However, if it scales like $N^\sigma$, in large enough systems the disease will not spread through the network in realistic timescales. 
We can address this question for the SI-model ($\gamma=0$) in an elegant fashion by establishing an analogy to the shortest path problem in disordered networks~\cite{ShlomoDiorderPRL1, ShlomoDiorderPRL2, PhysRevE.70.046133, VanMieghemDisorder1, VanMieghemDisorder2}, which are generated by assigning weights $\rho$ independently to edges drawn from some distribution $f(\rho)$. In~\cite{ShlomoDiorderPRL1} the authors have shown that the shortest path length from node $v_i$ to node $v_j$ in these weighted networks scales differently with the system size depending on $f(\rho)$, and distinguish between strong and weak disorder. 
The shortest path length is defined as $\mathit{d}(v_i,v_j) = \min_{\chi_{ij}} \sum_{(k,l)\in\chi_{ij}} \rho_{k,l}$, where $\chi_{ij}$ is the set of all possible paths.
For the example of Erdos-Renyi networks, in the case of strong disorder the shortest path length scales as $\left<d\right> \propto N^{1/3}$, and for weak disorder as $\left<d\right> \propto \log N$. 
The temporal distance we have defined earlier is equivalent to the shortest paths on weighted graphs.  
Therefore, we can use this analogy to study the scaling of the average propagation time with the system size. 
From~\cite{ShlomoDiorderPRL1} it is know that if $f(\rho)$ is taken as the exponential distribution, only weak disorder can occur, and the average distance scales with the logarithm of the systems size for Erdos-Renyi networks. 
This means that for the memory-less spreading process, where $\psi(\cdot)$ is the exponential distribution, the average propagation time scales with the logarithm of the system size and diseases spread quickly in systems of any size. 
However, if the inter-event distribution is for instance given by a lognormal distribution (non-Markovian case), we know from~~\cite{ShlomoDiorderPRL1} that strong disorder can occur such that the average distance and hence the average propagation time scales like $\left<d\right> \propto N^{1/3}$. In this case, the finite time properties of the spreading process are drastically different to the non-Markovian case, as the average propagation time becomes very large as the size of the system increases, such that the disease will not spread globally in realistic timescales.
To conclude, although it has recently been shown that the infinite time steady states of non-Markovian processes can be mapped to Markovian ones~\cite{PhysRevLett.118.128301},
their finite time dynamics can be dramatically different.

\section*{Results}

\subsection*{Expected propagation time between node pairs}
\begin{figure*}[h]
\centering
\includegraphics[width=0.8\linewidth]{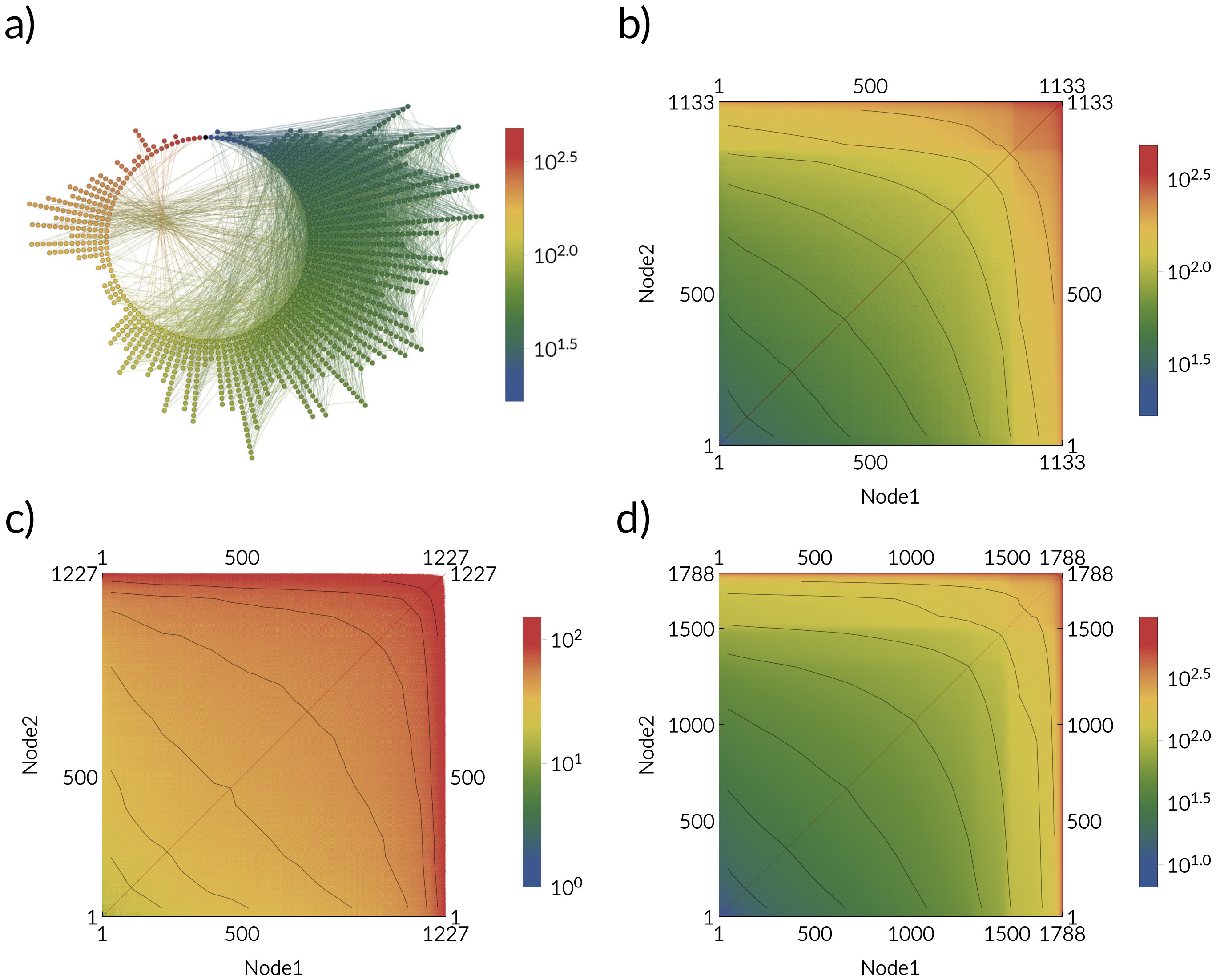}
\caption{ \textbf{The expected propagation time $D_{ij}$ on empirical networks}.
The average propagation time is calculated as the average shortest path distances in the ensemble of weighted networks (mapped dynamics).
\textbf{(a)} Nodes in the email network network are placed on a circle, where the angular coordinate represents the average propagation time from the source node (black) with index 500, increasing in the clockwise direction (from blue to red). 
The average propagation time $D_{ij}$ for: \textbf{(b)} the email network~\cite{email:dataset} ($\beta=0.01$), 
\textbf{(c)} the airport network ($\beta$ estimated from flux data see Supplementary information, Sec.~7), 
\textbf{(d)} the Petster network~\cite{petster:dataset} ($\beta=0.01$).
Nodes are ordered by their characteristic spreading timescale $\tau_i$, see Eq. \ref{eq:spreadingtimescale}) with condition $\overline{N}=N/2$. 
\label{figSI}}
\end{figure*}

Starting from a given node, what is the expected time after which the disease reaches each other node in the network? The answer to this question is encoded in the ensemble of weighted networks. 
For SI processes ($\gamma = 0$), the expected propagation time is always finite and
can be estimated as the average over $n$ samples of weighted networks as $D_{ij} = 1/n \sum_{k=1}^n f(\mathcal{G}_k)$, where $f(\mathcal{G}_k)=d_{\mathcal{G}_k}(v_i,v_j)$ denotes the length of weighted shortest path between $v_i$ and $v_j$. Note, that for the SIR process, the expected propagation time is not finite i.e. it diverges.
In Fig.~\ref{figSI}(a), we show an explicit example from the perspective of a specific source node. The advantage of the temporal maps we have created is they encode this type of information between all possible node pairs. The result is shown in Fig.~\ref{figSI}(b-d) for three empirical networks: the email network~\cite{email:dataset}, the airport network~\cite{Brockmann}, and the Petser online social network~\cite{petster:dataset}.
The plots show the average temporal distance or expected propagation time $D_{ij}$ for each pair of nodes $i,j$, and therefore encompass information about the expected spreading dynamics.
In particular, we have ordered the nodes by their \textit{characteristic spreading timescale}, which we define as the expected time after which a given node infects $\overline{N}$ other nodes, $\tau_i$, as
\begin{equation}
\tau_i = t|_{\sum_{j\neq i} \Theta(t-D_{ij}) = \overline{N}},
\label{eq:spreadingtimescale}
\end{equation}
where $\Theta(\cdot)$ denotes the Heaviside step function, which is one when $t\geq D_{ij}$ and zero otherwise. 
A small value of $\tau_i$ means that node $i$ is of high importance or influence. This ordering of nodes displays the smooth gradient of the average temporal distance between pairs of nodes on all three empirical networks.



\begin{figure}[h]
\begin{center}
\includegraphics[width=0.7\linewidth]{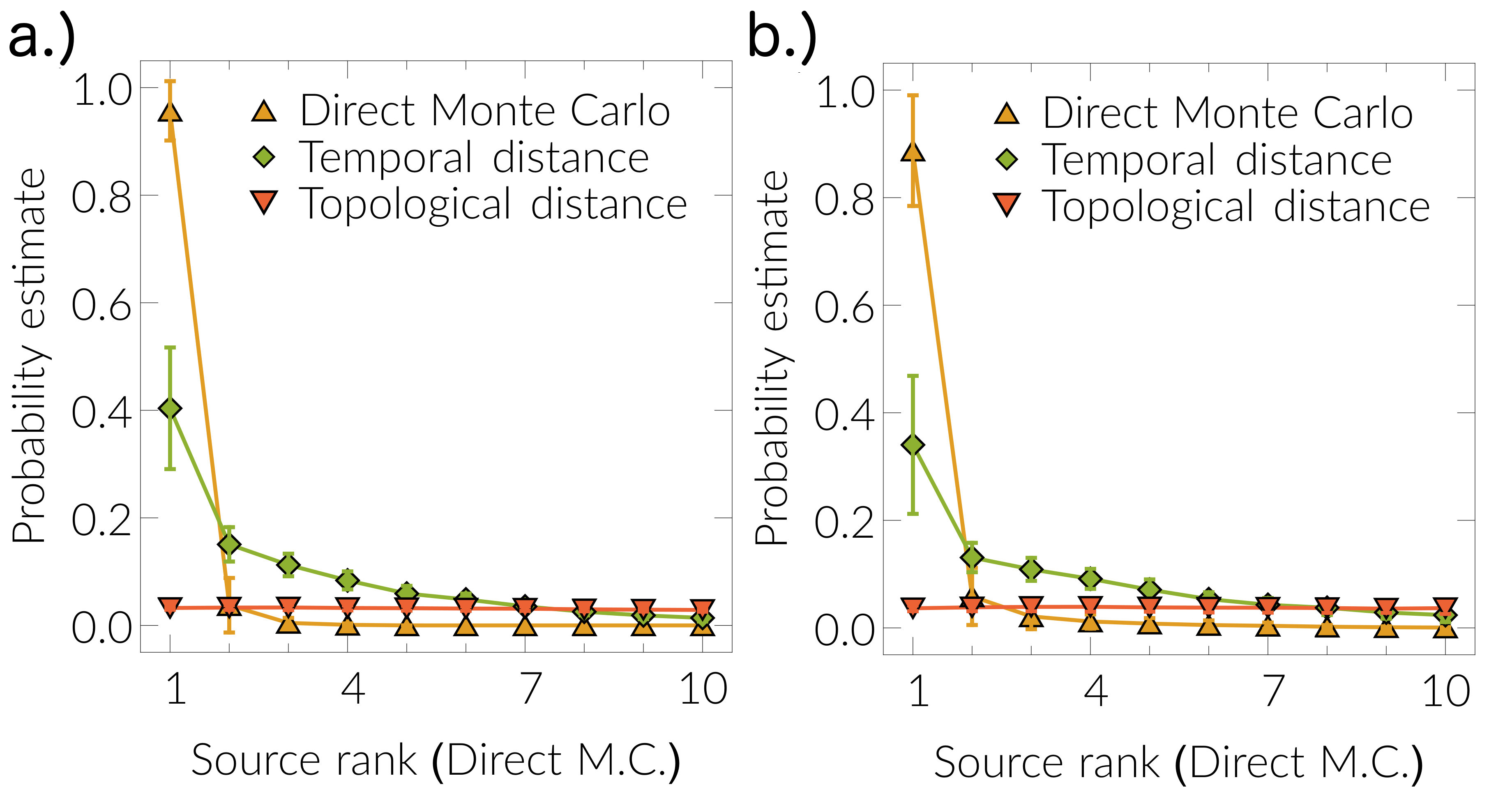}
  \end{center}
 \caption{
 \textbf{Source detection.} We 
 compare source detection performance for different methods for the SIR model: (\textbf{a}) $\beta=0.7, \gamma=0.3$ and (\textbf{b}) $\beta=0.7, \gamma=0.7$ at time $T=5$ (discrete time) on the 4-connected lattice (30x30 nodes). 
 For each observed realization, we rank nodes according to their estimated probability score according to the Direct Monte Carlo method \cite{AntulovFantulin2015}, in case of a tie we prioritize the node with a higher estimator according to the temporal distance method. 
 We compare different methods as explained in the text: Direct Monte Carlo source probability ranking, source detection ranking by topological distance \cite{Comin}, and the estimation from our mapping (temporal distance). 
 Results are averaged over $30$ observed realizations and error bars show one standard deviation, from top to bottom.
 \label{fig3}
 }
\end{figure}

\begin{figure}[h]
\begin{center}
\includegraphics[width=0.6\linewidth]{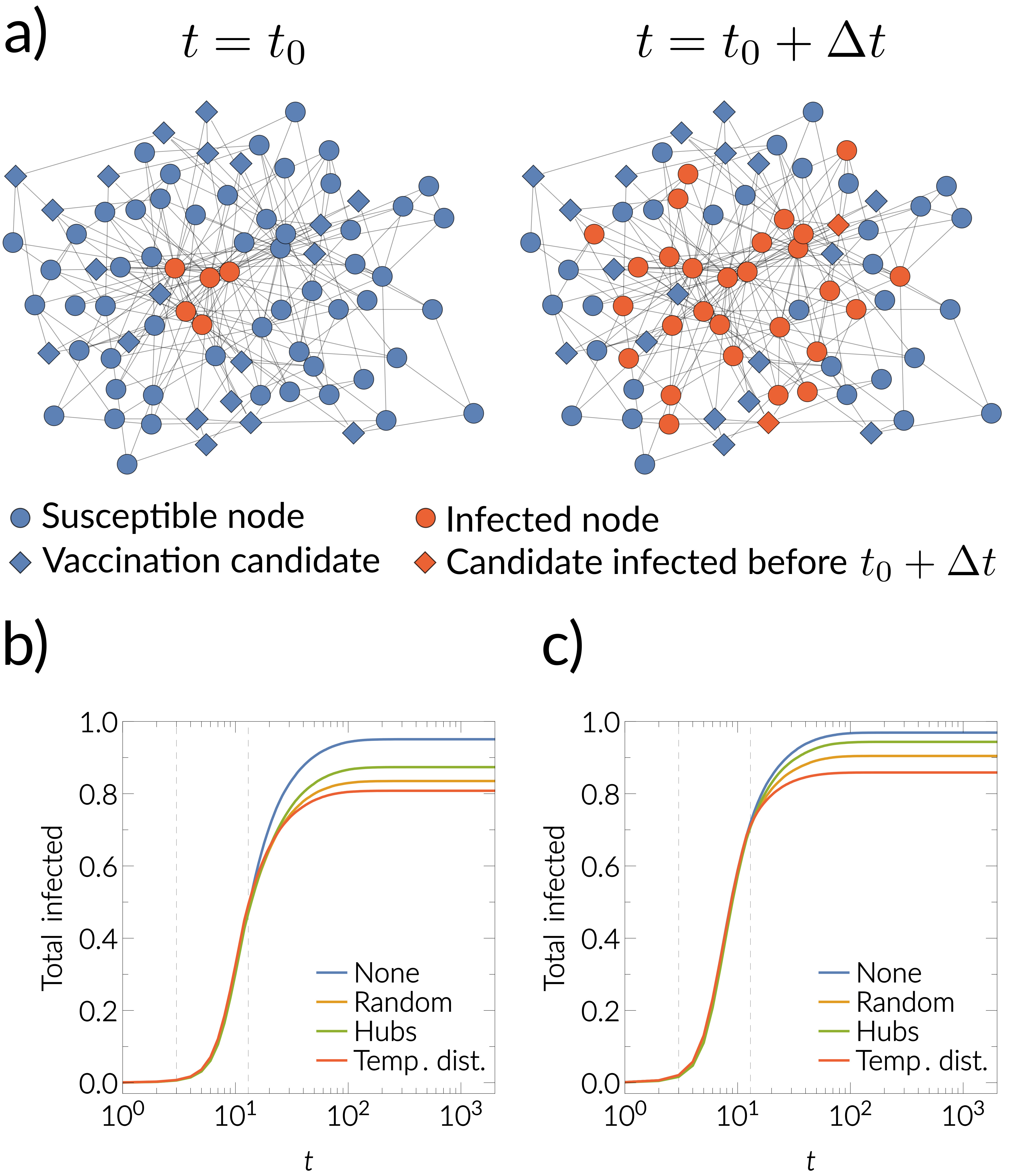}
\end{center}
\caption{ \textbf{Time critical vaccination.}
\textbf{(a)} 
Shows an illustration of the time critical vaccination process. 
Vaccination candidates are highlighted by the diamond symbols. Red note denote infected individuals, and blue nodes are susceptible.
In (\textbf{b}) and (\textbf{c}), we show the total number of infected nodes up to time $t$ for the different vaccination strategies described in the text ($m=0.2$ of the total number of nodes). The first dashed vertical line denotes $T_0=3$, and the second one $T_0 + \tau = 13$. In (\textbf{b}) for the discrete time SIR $\beta = 0.03, \gamma = 0.01$ and in  (\textbf{c}) for the discrete time SIR $\beta = 0.05, \gamma = 0.01$ with source node 10 in the Petster network.
\label{fig:vacc}
}
\end{figure}

\subsection*{Source detection}

For a given snapshot $r^*$ of a spreading process at time $t_0$ on a network, how does one detect the source node that generated the snapshot?
Source detection \cite{AntulovFantulin2015, Zaman1, pinto2012, DMP_0, BBP, Brockmann} is an important theoretical and practical problem in network science. It is at least as hard as any problem in \mbox{NP-complete} class~\cite{Zaman1, Zhai2015}, because it belongs to \mbox{\#P-complete} class~\cite{Provan1983}. We approach the problem by probabilistic kernel estimations \cite{AntulovFantulin2015, KernelEstimation} of the likelihood that source node generated the observed snapshot.
Our framework can be used to perform this task, as the ensemble of weighted networks encompasses information about different realizations of the spreading process from all possible sources. 

For the observed realization $r^*$, our source probability estimates for every node $v_i$ are constructed from ensemble averages $s_i \propto \sum_j^m f(\mathcal{G}_j)$.
From weighted network $\mathcal{G}_j$, we extract realization based on temporal distances and calculate the similarity $\varphi_j \in [0,1]$ to the observed realization $r^*$. 
Similarity $\varphi_j$ is the normalized number of equal corresponding states between two realizations.
Finally, we use the Gaussian kernel function 
\begin{equation}
f(\mathcal{G}_j) = exp(-(\varphi_j-1)^2/a^2)
\end{equation}
for probability estimations \cite{KernelEstimation, AntulovFantulin2015} (see Supplementary information, Sec.~4 for more details). 
We compare our source probability estimations with direct Monte Carlo approach~\cite{AntulovFantulin2015} and topological distance approach~\cite{Comin}. 
The direct Monte Carlo approach generates large number of simulations $n=10^6-10^8$, from each potential source node $v_i$, and source probability estimation is $s_i \propto n_i$, where $n_i$ is the number of simulations that match the observed realization $r^*$.
The topological distance approach calculates the average topological distance $\left\langle d_i \right\rangle$ from each potential source $v_i$ to other infected nodes in the observed realization $r^*$ and source estimation is $s_i \propto \left\langle d_i \right\rangle^{-1}$. 
The result is shown in Fig.~\ref{fig3}, where source ranked probability estimates for different methods are shown. Indeed, the method based on the ensemble of weighted networks ($n=10^5$ instances) performs better than a topological distance estimation~\cite{Comin} and correctly assigns a high probability to the source candidate identified by the direct Monte Carlo method~\cite{AntulovFantulin2015}.

\subsection*{Time critical vaccination}
Now, let us consider the following vaccination scenario. We observe at time $t_0$ the outbreak of a disease and we have a certain amount, $m$, of vaccines, see Fig.~\ref{fig:vacc}(a). We can vaccinate $m$ individuals instantaneously, but the vaccines will only be effective after some time $\Delta t$ (a vaccinated individual can get infected before $t_0+\Delta t$, and in this case this vaccine portion is wasted, see Fig.~\ref{fig:vacc}(b).
Who shall we vaccinate? 
Our framework can be used to improve the vaccination strategy. The idea is that, at time $t_0$, we vaccinate only individuals who---with high probability---will not get infected before $t_0+\Delta t$.
Therefore, we estimate the probability that a node $i$ will not be infected before $t_0+\Delta t$ as 
\begin{equation}
\tilde{p}_i = 1/n \sum_{k=1}^n \Theta(d_{\mathcal{G}_k}(v_\theta,v_i) - t_0-\Delta t), 
\end{equation}
where $\theta$ denotes the source node and $\Theta(\cdot)$ is one when $d_{\mathcal{G}_k}(v_\theta,v_i) \geq t_0+\Delta t$ and zero otherwise.
We vaccinate nodes proportional to $\tilde{p}$.
In Fig.~\ref{fig:vacc}(b-c) we show using empirical Petster network that this procedure performs better than randomly choosing from all susceptible nodes at $t_0$. Interestingly, vaccinating proportional to degree (hubs strategy) performs even worse than the fully random strategy, since the hubs usually are infected earlier, which increases the chance to waste vaccine portions.

\section*{Discussion}

We have mapped spreading dynamics on networks to an ensemble of weighted networks. This mapping allows us to define a temporal distance as shortest paths between nodes on an ensemble of weighted networks. 
We concentrate on the stochastic formulation of the generalized continuous and discrete time Susceptible Infected Recovered (SIR) spreading dynamics without memory (exponential inter-event distribution) and with memory (arbitrary inter-event distributions).  


The problem of shortest path with the stochastic independent and identically distributed weights (i.i.d.) has been extensively researched in theoretical computer science \cite{Valiant1979, Kulkarni1986, Corea1993, Peer2007}, probability theory \cite{VANDERHOFSTAD2006, Bhamidi2015} and statistical physics community \cite{ShlomoDiorderPRL1, ShlomoDiorderPRL2, PhysRevE.70.046133, VanMieghemDisorder1, VanMieghemDisorder2}.
However, the general SIR dynamics is exactly represented with networks where the edge weights are conditionally independent random variables and only in special case of spreading they become independent variables e.g. SI dynamics ($\gamma=0$) or SIR dynamics with fixed recovery time ($\psi(t)=\delta(t-T)$).
Thus, our mapping to the ensemble of weighted networks, provides additional dynamical interpretation of the shortest paths problem in disordered networks. Previously, only the Independent Cascade spreading models (edge weights are i.i.d.) were connected to the problem of shortest paths \cite{GomezRodriguez2016} on networks, but with no connections to disordered networks or bond percolation literature.

In contrast to message passing, our framework is applicable for arbitrary network structures without neglecting loops in the network structure.
Furthermore, our method takes into account dynamical correlations between node states overcoming the assumptions of mean-field like approximations \cite{Sharkey2011,  Castellano10}.
In contrast to the well known historical Gillespie or kinetic Monte Carlo methods \cite{Gillespie, dynamicMC, kinteicMC, BogunaNonMarkovian, TemporalNetGillespie, FastSIR, AntulovFantulin2015},
we do not have to specify initial conditions upfront and
we can sample new realizations from the previous ones by making local random perturbations on the weighted networks.
We have shown that for some non-Markovian processes, the average propagation time can scale as a polynomial with the system size, even if the underlying network is small world. In stark contrast to the Markovian case, this means that for large enough systems the disease does not spread globally in a realistic timescale.
We have shown that our framework allows for direct applications like ranking of nodes according to their characteristic spreading timescale, source detection and time-critical vaccination. Combining our mapping framework with existing vaccination strategies is likely to further improve the results, especially when time is a crucial factor. 

Note that previous studies~\cite{Brockmann, GAUTREAU2008509, COLIZZA2008450, Lawyer2016} for estimating the spreading arrival times were concentrated on a global disease spread with different approximations and theoretical understanding of effective distances~\cite{effDistBrockmann} for metapopulation models. 
However, the exact mathematical equivalences for mapping generalized spreading dynamics 
at a microscopic level were still lacking.
Furthermore, in a limit of infinite process time, we establish the connection with bond percolation theory. 
Previously the outcome in infinite time of the SIR process have been successfully mapped to the percolation \cite{mollison1977spatial, GrassbergerPercolation, NewmanPercolation, EpiPercolationNet} but not for the finite time.  
Contrary to previous state-of-the-art source inference studies \cite{AntulovFantulin2015, BBP, DMP_0}, estimations with our method require same computational effort to sample realizations at an arbitrary point in time $t$ not only for discrete time but also for continuous time dynamics.

\section*{Acknowledgments}
The authors would like to thank to prof. D. Helbing, A. Lancic, T. Lipic, O. Woolley and K. Caleb for useful comments, constructive feedback and discussions. 
The work of N.A.F. has been funded by the EU Horizon 2020 SoBigData project under grant agreement No. 654024. and in part by the EU Horizon 2020 CIMPLEX project under grant agreement No. 641191.
The work of D.T. has been funded by 
the Croatian Science Foundation IP-2013-11-9623, "Machine learning algorithms for insightful analysis of complex data structures".
K-K. K. acknowledges support by the ERC Grant “Momentum” (324247).

\section*{Author contributions}
All authors contributed to the writing and editing of the manuscript, proofreading and analyses. D.T. contributed to defining, formalizing and generalizing the mapping, relation to bond percolation and applications,
K.K.K. contributed to defining temporal maps, time critical vaccination and connection to disordered networks and
N.A.F. contributed to probabilistic modelling, source inference, algorithmic contributions and lead the research.

\section*{Additional information}
Dataset availability: All data that we used in our analysis are freely available~\cite{petster:dataset, email:dataset, BrockWAN}.\\
Supplementary information accompanies this paper.\\
The authors declare no competing interests.\\
Main code available at: https://github.com/ninoaf/SpreadingMapping \\

\bibliography{networks.bib}

\end{document}